\documentclass{article}
\pdfoutput=1

\usepackage{arxiv}

\usepackage[utf8]{inputenc} % allow utf-8 input
\usepackage[T1]{fontenc}    % use 8-bit T1 fonts
\usepackage{hyperref}       % hyperlinks
\usepackage{url}            % simple URL typesetting
\usepackage{booktabs}       % professional-quality tables
\usepackage{amsfonts}       % blackboard math symbols
\usepackage{nicefrac}       % compact symbols for 1/2, etc.
\usepackage{microtype}      % microtypography
\usepackage{lipsum}		    % Can be removed after putting your text content

\RequirePackage[]{natbib}

\usepackage[american]{babel}
\usepackage{csquotes}
\usepackage{amsmath}
\usepackage{todonotes}
\usepackage{tikz}
\usepackage{float}
\usepackage{lineno}
\usepackage{multirow}
\usepackage{adjustbox}
\usepackage{amsthm}
\usepackage{amssymb}
\usepackage{graphicx}
\usepackage[figuresright]{rotating}
\usepackage{comment}

\def\BF{\textsc{BF}}

\title{When Evidence and Significance Collide}
\author{
    František Bartoš                    \\
	Department of Psychological Methods \\
	University of Amsterdam             \\
	Noord-Holland, The Netherlands      \\
	\And
	Samuel Pawel                        \\
	Department of Biostatistics         \\
	University of Zurich                \\
	Zurich, Switzerland                 \\
	\And
	Eric-Jan Wagenmakers                \\
	Department of Psychological Methods \\
	University of Amsterdam             \\
	Noord-Holland, The Netherlands      \\
}

\renewcommand{\shorttitle}{Significance vs. Evidence}

\begin{document}
\maketitle

%\begin{abstract}
%No Abstract?
%\end{abstract}


Null hypothesis statistical significance testing (NHST) is the dominant approach for evaluating results from randomized controlled trials. Whereas NHST comes with long-run error rate guarantees, its main inferential tool --the \emph{p}-value-- is only an indirect measure of evidence against the null hypothesis. The main reason is that the \emph{p}-value is based on the assumption the null hypothesis is true, whereas the likelihood of the data under any alternative hypothesis is ignored.  
If the goal is to quantify how much evidence the data provide for or against the null hypothesis it is unavoidable that an alternative hypothesis be specified
\citep{Goodman1988}. Paradoxes arise when researchers interpret $p$-values as evidence.
For instance, results that are surprising under the null may be equally surprising under a plausible alternative hypothesis, such that a $p=.045$ result (`reject the null') does not make the null any less plausible than it was before. Hence, $p$-values have been argued to overestimate the evidence against the null hypothesis.
Conversely, it can be the case that statistically non-significant results (i.e., $p>.05)$ nevertheless provide some evidence in favor of the alternative hypothesis. It is therefore crucial for researchers to know when statistical significance and evidence collide, and this requires that a direct measure of evidence is computed and presented alongside the traditional $p$-value. 

In order to quantify evidence we need to know how likely the data are under a null hypothesis $\mathcal{H}_0$ relative to an alternative hypothesis $\mathcal{H}_1$. Within the framework of Bayesian hypothesis testing the relative model likelihood is known as the Bayes factor:
\begin{equation*}
   \underbrace{ \frac{p(\mathcal{H}_1 \mid \text{data})}{p(\mathcal{H}_0 \mid \text{data})}}_{\substack{\text{Posterior odds}}} \, = \, \underbrace{ \frac{p(\text{data} \mid \mathcal{H}_1)}{p(\text{data} \mid \mathcal{H}_0)} }_{\substack{\text{Bayes factor}~ \text{BF}_{10}}} \, \times \, \underbrace{ \frac{p(\mathcal{H}_1)}{p(\mathcal{H}_0)}}_{\substack{\text{Prior odds}}}.
\end{equation*}
The Bayes factor equals the degree to which the data mandate a change from prior to posterior odds; hence, it dictates how researchers should rationally update their beliefs based on the observed data. 
Alternatively, the Bayes factor can be interpreted as the degree to which one hypothesis outpredicted the other for the observed data -- hence it is not necessary to regard one of the hypotheses as ``true'' and the
other as ``false'' \citep{kass1995bayes}.

Bayes factors are not without criticisms, the most common being their dependence on the exact specification of the compared hypotheses. Whereas there is usually no dispute about the null hypothesis (usually taken to be $\mathcal{H}_0 \colon \text{effect} = 0$), specification of the alternative hypothesis is more controversial. Consequently, a plethora of approaches have been proposed to address this issue, such as sensitivity analysis, prior elicitation, objective Bayesian methodology, and reverse-Bayes analysis. In the following we will illustrate these approaches on two recently published clinical trials.

% Although an absolute ``objectivity'' is unattainable, we can still provide multitude of valuable insights into the evidence obtained from any particular study. First, we can specify a range of plausible alternative hypotheses and assess the evidence for each one of them (a procedure called prior sensitivity analysis). If we do not find compelling evidence for even a single plausible alternative hypothesis, we can concluded that the data are not persuasive. However, if most of the sensible alternative hypothesis are well-likely under the observed data, we can rest assured that our conclusions are sufficiently supported by the data). A similar argument can be also made on the grounds of Edwards--Jeffreys Bayes factor bound, $\text{BF}_{10}^\text{EJ}$ that specifies range of evidence obtainable by the most precise (i.e., an a priory guess of the to-be-observed outcome) versus a default (unit information) hypothesis \citep{wagenmakers2021history}. Second, we can find the most plausible alternative hypotheses that would yield compelling evidence (a procedure called reverse-Bayes methods; \citealp[e.g.,][]{held2021reverse}). If we do not find a reasonable alternative hypothesis that would provide compelling evidence, we can, again, conclude that the data are not persuasive.

\begin{figure}[ht]
    \centering
    \includegraphics[width=1\textwidth]{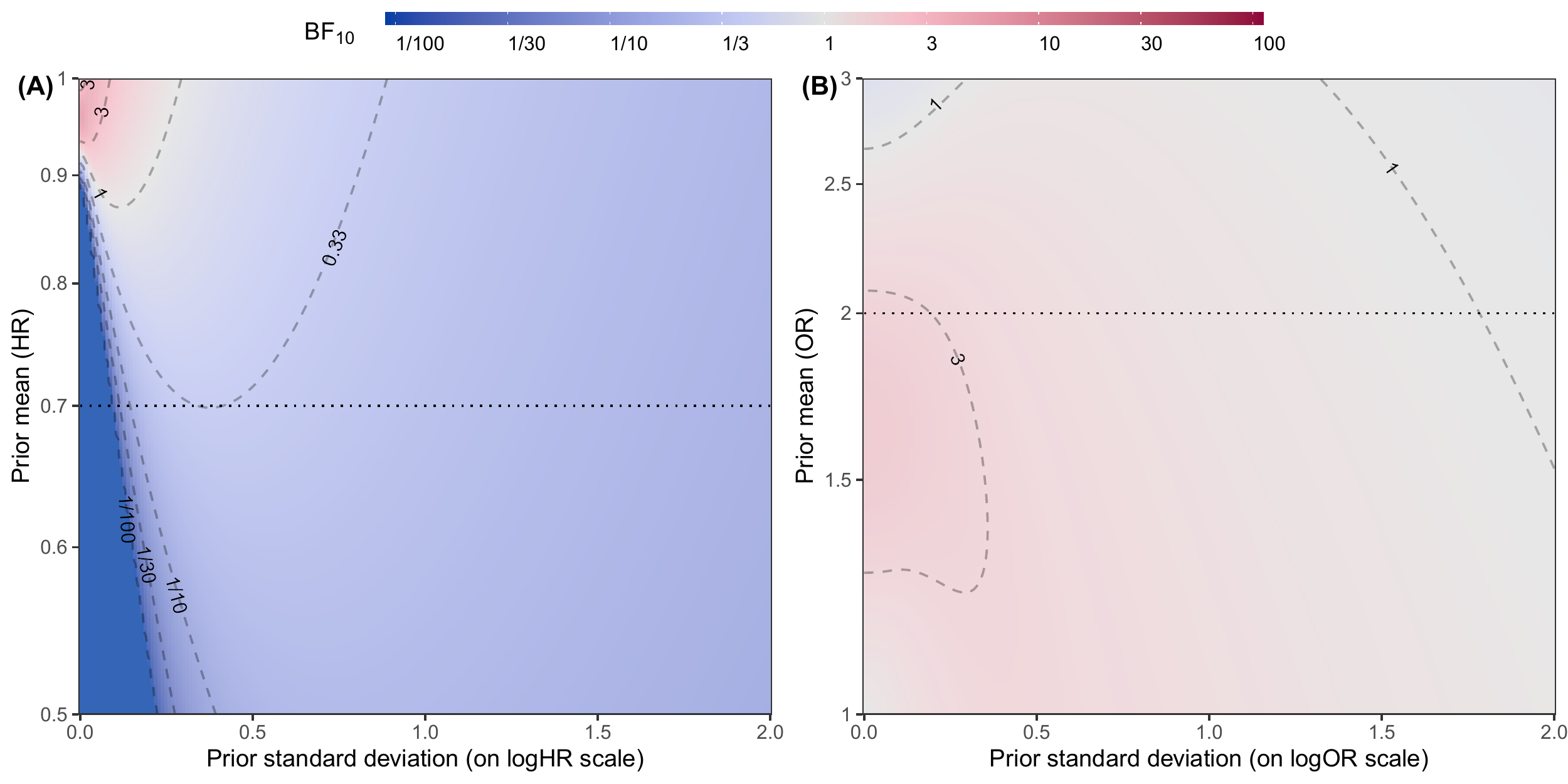}
    \caption{Bayes factor sensitivity analysis for two example data sets. The left panel shows results of \citet{svennberg2021clinical} with $\text{HR} = 0.96$, 95\% CI from 0.92 to 1, and $p = 0.045$. The right panel shows results of \citet{belohlavek2022effect} with $\text{OR} = 1.63$, 95\% CI from 0.93 to 2.85, and $p = 0.09$. The Bayes factor against the null hypothesis is shown for different prior means and standard deviations of the effect size under the alternative hypothesis. All calculations assume normal likelihood for the natural logarithm of the effect estimates and normal prior distribution for the underlying (log) effect size truncated on beneficial treatment effects (i.e., logHR < 0 and logOR > 0). For prior with standard deviation of zero the prior becomes a point mass at the location of the prior mean. Regions with Bayes factors smaller than $\nicefrac{1}{100}$ are coded with the same color. The horizontal dotted line depicts the effect size value used in the power analysis of the respective original trial.}
    \label{fig:sensitivity_analysis}
\end{figure}

The first example concerns a randomized controlled trial in which \citet{svennberg2021clinical} assessed the effectiveness of systematic screening on atrial fibrillation (a leading cause of ischaemic stroke) in an elderly population. Based on the observed data ($\text{HR} = 0.96$, 95\% CI from 0.92 to 1, $p = 0.045$), the authors concluded that ``Screening for atrial fibrillation showed a small net benefit compared with standard of care, indicating that screening is safe and beneficial in older populations'' (p. 1498). The left panel of Figure~\ref{fig:sensitivity_analysis} shows the Bayes factor against the null hypothesis $\mathcal{H}_0\colon\text{logHR} = 0$ for a range of different prior distribution of the $\text{logHR}$ under the alternative. Such a sensitivity analysis allows researchers to judge the extent to which the evidence is robust to changes in the specification of the alternative hypothesis. It shows there is at most moderate evidence in favor of a beneficial treatment effect (topleft corner, maximum $\text{BF}_{10} =$ 7.46), despite its statistical significance. Furthermore, for the alternative hypothesis that the researchers considered most plausible a priori (HR = 0.7, dotted horizontal line) and therefore used for the sample size calculation, the Bayes factor suggests decisive evidence in favor of the null hypothesis ($\BF_{10} < \nicefrac{1}{1000}$). Other reasonable alternative hypotheses -- small protective effects with some degree of uncertainty about the protectiveness, e.g., from HR = 0.7 to HR = 1 -- receive at most weak support from the data.

Another option is to use an objective Bayesian approach where the specification of the alternative hypotheses is based on generally applicable, problem-independent rules. For instance, researchers may specify a ``default'' unit-information prior that is centered on no effect with variance corresponding to a single observation. This approach also results in moderate evidence for the null hypothesis, $\text{BF}_{10} = \nicefrac{1}{6.73} = 0.149$. %We may go one step further and use a reverse-Bayes approach \citep{held2021reverse} asking the question ``which prior is required to obtain convincing evidence for the protectiveness of the treatment''. %For example, to obtain a Bayes factor favoring the alternative over the null hypothesis at least 5 times, unrealistic prior distributions which are tightly concentrated around the observed effect estimate are required. 
Taken together, our results illustrate that the trial fails to produce compelling evidence for a beneficial effect, despite the fact that the $p$-value was statistically significant.

The second example concerns a randomized controlled trial in which \citet{belohlavek2022effect} assessed whether or not an early invasive approach in adults with refractory extracorporeal cardiopulmonary resuscitation improves neurologically favorable survival. The trial led to an effect estimate of $\text{OR} =$ 1.63 (95\% CI from 0.93 to 2.85, $p =$ 0.09), and based on this result the authors conclude that the intervention ``did not significantly improve survival with neurologically favorable outcome'' with the caveat that ``the trial was possibly underpowered to detect a clinically relevant difference'' (p. 737). Despite this negative conclusion, the right panel of Figure~\ref{fig:sensitivity_analysis} paints a picture that differs from the previous example; while the maximum evidence against the null hypothesis (maximum $\text{BF}_{10} =$ 4.31) is slightly smaller, the Bayes factor indicates weak to moderate support for a wide range of alternative hypotheses, and the default unit-information prior hypothesis shows little evidence against the null ($\text{BF}_{10} = 1.13$).

We may go one step further and use a reverse-Bayes approach \citep{held2021reverse} to address the question ``which prior is required to obtain evidence for the protectiveness of the treatment''. In the second example, a broad range of prior distributions -- including small to large effect sizes -- results in a Bayes factor that favors the alternative over the null hypothesis ($\text{BF}_{10} >$ 1). In the first example, the prior range is much more limited. The second trial thus presents some evidence for plausible effect sizes, regardless of its non-significance, although the extent of this evidence is not compelling.

In sum, we propose that the standard measure of statistical significance (i.e., the $p$-value) is supplemented by the standard measure of statistical evidence (i.e., the Bayes factor). This is especially important in medicine, where a confusion between significance and evidence may cause researchers to recommend a treatment that is actually contraindicated by the evidence, or to abandon a treatment that is actually supported by the evidence. Bayes factors formalize what many researchers wish to do when assessing evidence through an intuitive joint evaluation of $p$-values, effect sizes, power, and sample sizes. Several tools exist for assessing the impact of different alternative hypotheses on the resulting Bayes factor. Collectively, these approaches allow researchers to draw sensible inferences that reflect the available evidence from their data.

\section*{Acknowledgement}

This work was supported in part by a Vici grant from the Netherlands Organization of Scientific Research (NWO; 016.Vici.170.083)
to Eric-Jan Wagenmakers, and a 
Swiss National Science Foundation mobility grant (part of 189295) to Samuel Pawel.

\section*{Supplementary material}

The analysis script is available at \url{https://osf.io/hvmkc/}.

\bibliographystyle{biometrika}
\bibliography{manuscript.bib}

\end{document}